%% file: ms_ieee.tex
\documentclass[letterpaper, 10 pt, conference]{ieeeconf}
\input{preamble.tex}

\title{\vspace{0.4cm} Control Node Selection Algorithm for Nonlinear Dynamic Networks}

\author{Aleksandar Habe$\text{r}^1$,  Sebastian A. Nugroh$\text{o}^{2}$, Patricio Torre$\text{s}^{3}$, and Ahmad F. Tah$\text{a}^{2}$\vspace{-0.5cm}
	\thanks{
		$^1$Department of Engineering and Environmental Science, City University of New York, College of Staten Island, 2800 Victory Blvd, New York, NY 10314, USA. $^{2}$Department of Electrical and Computer Engineering, The University of Texas at San Antonio, 1 UTSA Circle, San Antonio, TX 78249, USA. $^{3}$Center for Mathematical Modeling, Department of Mathematical Engineering, University of Chile,  Santiago, Chile.		
			Emails: aleksandar.haber@csi.cuny.edu, sebastian.nugroho@my.utsa.edu, ahmad.taha@utsa.edu, pattorre@gmail.com. This work is supported by the PSC-CUNY Award A (61303-00 49), the PSC-CUNY Award A (62267-00 50), and National Science Foundation Grants CMMI-1728629 and CMMI-1917164.}
}

\begin{document}

\maketitle

\setlength{\abovedisplayskip}{3.5pt}
\setlength{\belowdisplayskip}{3.5pt}
\setlength{\abovedisplayshortskip}{3.1pt}
\setlength{\belowdisplayshortskip}{3.1pt}

\newdimen\origiwspc%
\newdimen\origiwstr%
\origiwspc=\fontdimen2\font
\origiwstr=\fontdimen3\font

\fontdimen2\font=0.64ex

\begin{abstract}The coupled problems of selecting control nodes and designing control actions for nonlinear network dynamics are fundamental scientific problems with applications in many diverse fields. These problems are thoroughly studied for linear dynamics; however, in spite of a number of open research questions, methods for nonlinear network dynamics are less developed. As observed by various studies, the prevailing graph-based controllability approaches for selecting control nodes might result in significantly suboptimal control performance for nonlinear dynamics. Herein we present a new, intuitive, and simple method for simultaneous control node selection and control sequence design for complex networks with nonlinear dynamics. The method is developed by incorporating the control node selection problem into an open-loop predictive control cost function and by solving the resulting mixed-integer optimization problem using a mesh adaptive direct search method. The developed framework is numerically robust and can deal with stiff networks, networks with non-smooth dynamics, as well as with control and actuator constraints. Good numerical performance of the method is demonstrated by testing it on prototypical Duffing oscillator and associative memory networks. The developed codes that can easily be adapted to models of other complex systems are available online.
\end{abstract}

%

\section{Introduction}
The fundamental problems of controlling and estimating states of nonlinear network dynamics and systems  {appear in a large variety of} engineering and scientific disciplines. These problems are crucial for the design, safe operation, and analysis of power systems, elastic structures, electrical circuits, traffic, communication, chemical reaction, ecological, and biological networks~\cite{liu2011controllability,liu2013observability,cornelius2013realistic,angulo2019theoretical,haber2014subspace,haber2018sparsity}. 
The control problem for nonlinear networks consists of two subproblems. The first subproblem, referred to as the \textit{control node selection problem}, is to select a subset of control nodes such that the network is controllable. Once the control nodes have been selected, the second subproblem, referred to as the \textit{control design problem}, is to design control actions that will achieve the desired system performance. Closely related problems to these two subproblems are  \textit{sensor selection and observer design problems} that deal with state estimation of nonlinear network dynamics. The necessity for selecting control nodes originates from the fact that it is often expensive to control all nodes in the network or it is physically impossible to install actuators on every node.

 A large number of recent approaches for sensor and control node selection rely on graph-theoretic methods that are widely popularized and revived in~\cite{liu2011controllability,liu2013observability}.
These approaches and a large number of follow-up contributions rely on the main results of control theory for systems with graph structure~\cite{reinschke1988multivariable}. Although such approaches can provide us with some insights and preliminary solutions of control and sensor selection problems, a number of authors have observed and analyzed several shortcomings of graph-based approaches~\cite{pasqualetti2014controllability,haber2017state,summers2015submodularity,letellier2018nonlinear,aguirre2018structural}. Among several limitations, the main limitation that is relevant for this work is that, in some cases, graph-based approaches might result in far from optimal control solutions. As possible remedies to these limitations, the authors in  \cite{summers2015submodularity,pasqualetti2014controllability} propose methods to optimally select control (sensor) nodes for linear networks by optimizing Gramian-based controllability (observability) metrics.  However, such methods are designed for linear network dynamics. 

 Sensor selection and state estimation problems for nonlinear networks have been considered in~\cite{haber2017state,qi2014optimal}. Despite the fact that control and estimation problems are dual, the generalization of the approach presented in~\cite{haber2017state} for control node selection is {not} straightforward.  {In principle, empirical Gramian-based approaches used in~\cite{qi2014optimal} and summarized in~\cite{haber2017state}, can also be used for control node selection.} The drawback of these approaches is that the computation of empirical controllability Gramians is computationally prohibitive even for small-sized networks. Recently, control node selection algorithms for linear systems have been proposed in~\cite{nugroho2019algorithms,taha2018time,chang2018co,hao2019linear,taylor2016allocating}. The applicability of these methods to nonlinear systems has to be theoretically and numerically investigated.

Recently, a new approach for sensor selection and observer design for nonlinear systems has been presented in~\cite{nugroho2019sensor}. The potential of using this method for control node selection still has to be investigated, especially for networks with stiff dynamics that are {ubiquitous in real-life applications}. Traditional control-theoretic approaches for actuator and sensor placement for linear systems have been summarized~\cite{van2001review}. Sensor/actuator placement problems for the system dynamics described by partial differential equations have been considered in~\cite{lou2003optimal,morris2010linear,edalatzadeh2019optimal2,edalatzadeh2019optimal}. To the best of our knowledge, most of the approaches for control node selection overlook the fact that often in practice, the network dynamics can be stiff~\cite{haber2017state} or even non-smooth. Finally, control node selection and control design problems are usually treated separately which might result in far from optimal performance. 

To address the limitations of the approaches discussed above, we develop a novel control node selection method. The basic idea of our approach is to simultaneously compute an optimal selection of control nodes and control actions. This is achieved by integrating the control node selection problem into an open-loop predictive control cost function. As a result, we obtain a Mixed-Integer Nonlinear Optimization (MINO) problem whose solutions are optimal locations of control nodes and control actions. Despite the fact that in the general case the resulting MINO problem is nonconvex and NP-hard, by performing extensive numerical experiments, we show that such a problem can be effectively solved using the Mesh Adaptive Direct Search (MADS) algorithm~\cite{le2011algorithm}. The main advantage of this solution process over other solution methods relying on variations of a branch and bound method~\cite{belotti2013mixed} is in its implementation simplicity and generality. Namely, the used approach is applicable to a broad class of nonlinear network dynamics and it does not rely on convexification or linearization procedures that are often case dependent. Besides this, the developed approach is numerically robust and it can easily handle stiff or non-smooth dynamics as well as actuator and various control constraints. We test the developed approach on models of prototypical Duffing oscillators networks, resembling models of many complex systems, as well as on associative memory networks, representing memory models. Good numerical performance is confirmed by testing the method against exhaustive search and random control node selection. The used codes are available online~\cite{haberControlCode2020}.

 The letter is organized as follows. In Section~\ref{nonlinearNetworks}, we introduce the class of networks considered in this letter and preliminaries. Next, in  Section~\ref{controlProblemFormulation} we formulate the MINO control problem and develop the solution method. In Section~\ref{numericalResults} and Section~\ref{conclusionsSection} we present numerical results and conclusions, respectively.

\section{System Descriptions and Preliminaries}
\label{nonlinearNetworks}

In this section, we describe the class of considered networks. We consider nonlinear networks composed of $N$ nodes. The network model with parametrized locations of control nodes has the following form
\begin{align}
\dot{\mathbf{x}}(t)=\mathbf{f}(\mathbf{x})+\mathbf{B}\big(\boldsymbol{\pi} \big) \mathbf{z}(t),
\label{parametrizedDynamics}
\end{align}
where $\boldsymbol{\pi}\in \{0,1\}^{N}$ is a binary parameterization vector variable encoding control node locations, $\mathbf{x}=\mathrm{col}\big(\mathbf{x}^{(1)},\mathbf{x}^{(2)}, \ldots, \mathbf{x}^{(N)} \big)\in \mathbb{R}^{Nn}$ is the network global state consisting of node states $x^{(i)}\in \mathbb{R}^{n}$ (the operator $\mathrm{col}(\cdot)$ stacks all vectors into a single vector), $\mathbf{f}(\mathbf{x}) : \mathbb{R}^{Nn} \rightarrow \mathbb{R}^{Nn}$ is a nonlinear function, and $\mathbf{z}\in \mathbb{R}^{N}$ is a parametrized control input allowing all nodes to be controlled by a scalar input. For simplicity we assume that the local control input is one-dimensional. For presentation clarity, we assume that the control input affinely influences the system dynamics. The developed method can be straightforwardly generalized to the case of nonlinear dependencies between the state dynamics and control inputs. The non-zero pattern of $\boldsymbol{\pi}$ determines the set of controlled nodes. The parameterization of the matrix $\mathbf{B}(\boldsymbol{\pi})\in \mathbb{R}^{Nn\times N}$ depends on how the control input affects the dynamics. In numerical models considered in this letter, we assume that the matrix $\mathbf{B}\big(\boldsymbol{\pi} \big)$ is parameterized as follows. For networks with $n=1$, we have  $\mathbf{B}\big(\boldsymbol{\pi} \big)=\text{diag}(\boldsymbol{\pi})$, where the notation $\text{diag}(\cdot)$ is used to define a  diagonal matrix with the vector $\boldsymbol{\pi}$ on the main diagonal. For  networks with $n>1$ we assume that the matrix $\mathbf{B}$ is a block diagonal matrix, with the $i$-th block $\mathbf{B}_{i}=\text{col}(0,0,\ldots, \pi_{i} )$. That is, the $i$-th column of $\mathbf{B}$ is zero if $\pi_i = 0$. This is equivalent to having the zero control action at node $i$. Under the constraint that $M$ nodes can be controlled, our goal is to determine the non-zero pattern of $\boldsymbol{\pi}$. Once this is completed, we can eliminate zero columns of $\mathbf{B}$ in order to obtain the matrix $\hat{\mathbf{B}}\in \mathbb{R}^{Nn\times M}$. This procedure produces the following model
\begin{align}
\dot{\mathbf{x}}(t)=\mathbf{f}(\mathbf{x})+\hat{\mathbf{B}}\mathbf{u}(t),
\label{systemNonlinearDynamics}
\end{align}
where $\mathbf{u}\in \mathbb{R}^{M}$ represents the control input vector formed from the entries of $\mathbf{z}$. A large number of complex networks, such as highway traffic, combustion dynamics, and epidemic outbreak networks can be expressed by \eqref{systemNonlinearDynamics}. Under the constraint that a predefined number $M$, $M\le N$, of nodes can be used for control, our goal is to determine the structure of matrix $\hat{\mathbf{B}}$ and the sequence of control inputs  $\mathbf{u}$ that jointly optimize a control performance that will be defined in the sequel. 

To simplify the method development, we represent the dynamics in the discrete-time domain~\cite{lars2011nonlinear}. The choice of a method used to discretize the dynamics depends on desired accuracy, available computational resources, and the degree of \textit{stiffness}. Networks with stiff dynamics are characterized by time constants of local nodes that significantly differ in magnitude. Typical examples of such networks are ubiquitously present chemical reaction networks and networks coupling different physical phenomena, such as reaction-diffusion systems. To accommodate various networks with different stiffness properties, herein we consider two distinct discretization approaches. In the case of non-stiff network dynamics where all nodes have time constants of similar magnitudes, we consider the {Forward Euler} (FE) discretization method~\cite{iserles2009first}. The discretized dynamics of network \eqref{parametrizedDynamics} has the following form
 	\begin{align}
 	 	\mathbf{x}_{k+1}=\mathbf{x}_{k}+h\big(\mathbf{f}(\mathbf{x}_{k})+\mathbf{B}\big(\boldsymbol{\pi} \big)\mathbf{z}_{k}  \big),
 	\label{forwardEulerParametrized}
 	\end{align}
 	where $h>0$ is a discretization step, $\mathbf{x}_{k}:=\mathbf{x}(kh)$  and $\mathbf{z}_{k}:=\mathbf{z}(kh)$ are the discrete global state and discrete global input, respectively, and $k=0,1,2,\ldots$, is a discrete-time instant. 
 	If $h$ is a relatively small number and the network is not stiff, the FE method is able to accurately approximate the continuous-time dynamics. Note that in \eqref{forwardEulerParametrized} the future state $\mathbf{x}_{k+1}$ \textit{explicitly} depends on the current state $\mathbf{x}_{k}$ and input $\mathbf{z}_{k}$. In the case of stiff network dynamics, implicit discretization methods are required.	To that end, we employ the {Trapezoidal Implicit (TI)} method~\cite{iserles2009first} to \eqref{parametrizedDynamics}, which results in
 	\begin{equation}
\mathbf{x}_{k}=\mathbf{x}_{k-1}+\frac{h}{2}\Big(\mathbf{f}(\mathbf{x}_{k})+ \mathbf{f}(\mathbf{x}_{k-1})+\mathbf{B} \big(\boldsymbol{\pi} \big) (\mathbf{z}_{k}+\mathbf{z}_{k-1}) \Big). \label{TIdiscretizationParametrized} 
 	\end{equation}
 	In a sharp contrast to \eqref{forwardEulerParametrized}, notice that the current state $\mathbf{x}_k$ in \eqref{TIdiscretizationParametrized} \textit{implicitly} depends on current and prior states and inputs. A solution of the optimization problem defined in the sequel consists of a repeated simulation of the discretized dynamics. The main computational disadvantage of the TI method over the FE method is that in every simulation step $k$, we need to solve the  nonlinear system of equations in~\eqref{TIdiscretizationParametrized}, resulting in $\mathcal{O}\big( n^3N^3 \big)$ computational complexity. On the other hand, the simulation of the FE discretized dynamics consists of forward propagation of~\eqref{forwardEulerParametrized}, {resulting in $\mathcal{O}\big( nN \big)$ computational complexity.}  However, the main advantage of using the TI dynamics is that we can handle a much broader class of nonlinear systems as well as larger discretization steps than in the case of the FE dynamics.
 	

\section{Simultaneous Control Node and Control Action Designs: Solution Approach}
\label{controlProblemFormulation}
In this section, we present the control node selection method that is developed by incorporating the control node selection problem into a control action design problem. In order not to blur the main ideas of this paper and for brevity, for control action design we use an open-loop predictive control framework. The method proposed in this paper can be applied to more general model-based control frameworks, such as model predictive control approach~\cite{lars2011nonlinear}, for which it is possible to stabilize the system around unstable desired state. For a given initial state $\mathbf{x}_{0}$, the control action design consists of finding the control input sequence $\mathbf{u}_{0:T}:=\mathrm{col}\big(\mathbf{u}_{0},\mathbf{u}_{1},\ldots, \mathbf{u}_{T} \big)$, $\mathbf{u}_{k}:=\mathbf{u}(kh)$, that will drive the network state to be as close as possible to the desired state, denoted by $\mathbf{x}_{D}$, within a discrete-time window of the length of $T$, while at the same time optimizing a control performance criterion (cost function). We consider the following cost function that is parametrized by control node locations and control inputs
\begin{align}
J\big(\mathbf{z}_{0:T},\boldsymbol{\pi} \big)= \sum_{i=1}^{T} \big(  \mathbf{x}_{D}-\mathbf{x}_{i}  \big)^{T}\mathbf{Q}_{i} \big(  \mathbf{x}_{D}-\mathbf{x}_{i}  \big),
\label{controlCostFunction}
\end{align}
where $\mathbf{z}_{0:T}:=\mathrm{col}\big(\mathbf{z}_{0},\mathbf{z}_{1},\ldots, \mathbf{z}_{T} \big)$ and $\mathbf{Q}_{i}\in \mathbb{R}^{Nn}$ are the weighting matrices. The control node selection problem has the following form 
\begin{subequations}\label{problemP}
{
\begin{align}
&(\textbf{P})\;\min_{\mathbf{z}_{0:T},\boldsymbol{\pi}}  \;\;\; J\big(\mathbf{z}_{0:T},\boldsymbol{\pi} \big),
\label{OptimizationProblem1} \\
&\hspace{-0.2cm}\text{subject to}\;\;
\mathbf{x}_{i}=\mathbf{w}\big(\mathbf{x}_{i},\mathbf{x}_{i-1},\mathbf{z}_{i},\mathbf{z}_{i-1},\boldsymbol{\pi}\big), \; \forall i,  \label{constraint1} \\
&\sum_{l=1}^{N} \pi_{l} \le M_{\text{max}} , \;\pi_{l}\in \{0,1 \}. \label{constraint2}  
\end{align}
}
\end{subequations}
where $\mathbf{z}_{0:T}:=\mathrm{col}\big(\mathbf{z}_{0},\mathbf{z}_{1},\ldots, \mathbf{z}_{T} \big)$, $\mathbf{Q}_{i}\in \mathbb{R}^{Nn}$ are weighting matrices, $\pi_{l}$ is the $l$-th entry of $\m \pi$, and $\mathbf{w}(\cdot)$ stands for discretized dynamics that is defined in \eqref{forwardEulerParametrized} or \eqref{TIdiscretizationParametrized}, depending on the used discretization method. For presentation clarity, we have only used essential constraints in the MINO problem \eqref{constraint1}. The proposed method can easily be generalized to include constraints on the control inputs and hard constraints on the difference between the final and desired states. To solve the MINO problem, we utilize the MADS algorithm (also known as NOMAD) that is implemented in the OPTI MATLAB toolbox~\cite{currie2012opti}. {This is a derivative-free optimization method, only requiring a procedure to evaluate the cost function and constraints. } Furthermore, we have chosen MADS (NOMAD) due to its MATLAB interface and its ability to integrate all the nonlinear MATLAB solvers that are necessary to simulate the system dynamics. 
\setlength{\floatsep}{5pt}
{
	\begin{algorithm}[t]
		\caption{Control Node Selection}\label{algorithm1}
		\DontPrintSemicolon
		\textbf{inputs:} $\mathbf{x}_{0}$, $\mathbf{x}_{D}$, $M_{\max}$, $T$, and $\m Q_{i},i=1,\ldots,T$. \;
		\textbf{initial solution:} Set $\m \pi^{(0)} = \m 1$ and solve the NLP
		\vspace{-0.1cm}
		\begin{subequations}\label{problem0} {
				\begin{flalign}
&\min_{\mathbf{z}_{0:T}}  \;\;\;\;\;\; J\big(\mathbf{z}_{0:T}, \boldsymbol{\pi}^{(0)} \big), 
\label{OptimizationProblem0} \\
&\hspace{-0.2cm}\text{subject to}\;
\mathbf{x}_{i}=\mathbf{w}\big(\mathbf{x}_{i},\mathbf{x}_{i-1},\mathbf{z}_{i},\mathbf{z}_{i-1},\m \pi^{(0)}\big),\;\forall i.  \label{constraint0} 
\end{flalign}}
		\end{subequations}
		\;
		\textbf{solve:} \textbf{P} in \eqref{problemP} with  $({\boldsymbol{\pi}}^{(0)},{\mathbf{z}}_{0:T}^{(0)})$ as an initial guess, where ${\mathbf{z}}_{0:T}^{(0)}$ is the solution of \eqref{problem0}. Let $\hat{\m \pi}$ and $\hat{\mathbf{z}}_{0:T}$ be the resulting optimal solutions.\;
		\textbf{construct:} The new reduced matrix $\hat{\mathbf{B}}$ where its columns correspond to the nonzero columns of ${\mathbf{B}}(\hat{\m \pi}) $\;
		\textbf{final solution:} Using  $\hat{\mathbf{B}}$ as a fixed variable solve 
		\begin{subequations}\label{problem011} {
				\begin{flalign}
&\min_{\mathbf{u}_{0:T}}  \;\;\;\;\;\;  J\big(\mathbf{u}_{0:T}\big), 
\label{OptimizationProblem011} \\
&\hspace{-0.2cm}\text{subject to}\;\; 
\mathbf{x}_{i}=\mathbf{g}\big(\mathbf{x}_{i},\mathbf{x}_{i-1},\mathbf{u}_{i},\mathbf{u}_{i-1},\hat{\mathbf{B}}),\;\forall i.  \label{constraint011} 
\end{flalign}}
		\end{subequations}
    where $\mathbf{g}(\cdot)$ in \eqref{constraint011} is obtained by substituting $\mathbf{B} \big(\boldsymbol{\pi} \big)$ by $\hat{\mathbf{B}}$, and $\mathbf{z}_{i}$ by $\mathbf{u}_{i}$ in $\mathbf{w}(\cdot)$. Let $\hat{\mathbf{u}}_{0:T}$ be the corresponding optimal solution.\;
		\textbf{output:} $\hat{\m \pi}$, $\hat{\mathbf{u}}_{0:T}$.\;
\end{algorithm}}

The proposed method is summarized in Algorithm \ref{algorithm1}. To generate an initial solution for \textbf{P}, in step 2, we use the full set of control nodes, i.e. $\m \pi^{(0)} = \m 1$. This yields a NonLinear Program (NLP) described in \eqref{problem0} whose solution is represented by $\mathbf{z}_{0:T}^{(0)}$. This NLP problem is solved using the quasi-Newton method implemented in the MATLAB function $\texttt{fminunc}$. Another option for this step would be to relax the integer constraints in \textbf{P} by $0\le \pi_{l} \le 1$, to solve the resulting NLP problem for both $\mathbf{z}_{0:T}$ and $\m \pi$, and to threshold the entries of $\m \pi$ to either $0$ or $1$. However, for the network models considered in this letter both approaches generate similar results. In step 3, using the initial solution guess $(\boldsymbol{\pi}^{(0)},\mathbf{z}_{0:T}^{(0)})$ we approximately solve \textbf{P} using the NOMAD solver. The solutions obtained from this step are denoted by $(\hat{\boldsymbol{\pi}},\hat{\mathbf{z}}_{0:T})$. In step 4 we construct the reduced matrix $\hat{\mathbf{B}}$ from non-zero columns of $\mathbf{B}\big(\hat{\boldsymbol{\pi}} \big)$, and  in step 5, we solve the NLP problem in \eqref{problem011} to compute $\hat{\mathbf{u}}_{0:T}^{(0)}$. The final outcomes are $\hat{\boldsymbol{\pi}}$ (set of nodes that need to be controlled) and $\hat{\mathbf{u}}_{0:T}$ (control sequence to be applied to these nodes).  We use a recursive approach for solving the MINO and NLP problems, see Chapter 10 in~\cite{lars2011nonlinear}. Consequently, the states are not considered as explicit optimization variables. 

For some particular cases of $\mathbf{f}(\cdot)$, our problem can potentially be solved using other MINO solvers and approaches, see for example~\cite{achterberg2009scip,belotti2013mixed}. However, in most cases, the MINO problem has to be convexified, linearised, or represented in an equivalent form, and this procedure is case dependent and might be highly non-trivial for general forms of the dynamics $\mathbf{f}(\cdot)$. Furthermore, the extensions of branch and bound methods summarized in~\cite{achterberg2009scip,belotti2013mixed} are developed for problems that are not constrained by system dynamics, and their applicability to our case when the problem is constrained by the system dynamics requires further theoretical and numerical investigations. Consequently, it is challenging to implement the methods of \cite{achterberg2009scip,belotti2013mixed} and to compare them with our approach. To perform the comparison, we use an alternative approach that is inspired by an idea for solving MINO problems arising in the design of time-domain-sparse control inputs for predictive control~\cite{sager2009reformulations,sager2011combinatorial,burger2019design}. The idea is to relax the integer constraints and to first solve the relaxed problem
\begin{align}
&\min_{\mathbf{z}_{0:T},\boldsymbol{\alpha}}  \;\;\; J\big(\mathbf{z}_{0:T},\boldsymbol{\alpha} \big),
\label{OptimizationProblem1RelaxedSager} \\
&\hspace{-0.2cm}\text{subject to}\;\;
\mathbf{x}_{i}=\mathbf{w}\big(\mathbf{x}_{i},\mathbf{x}_{i-1},\mathbf{z}_{i},\mathbf{z}_{i-1},\boldsymbol{\alpha}\big), \; \forall i, \;  \mathbf{0} \le \boldsymbol{\alpha} \le \mathbf{1}  \label{constraint1RelaxedSager2}. 
\end{align}
where $\boldsymbol{\alpha}\in \mathbb{R}^{N}$ is a relaxation of the vector $\boldsymbol{\pi}$. The problem \eqref{OptimizationProblem1RelaxedSager}-\eqref{constraint1RelaxedSager2} belongs to the class of NLPs and we solve it using the interior point method implemented in the MATLAB function \texttt{fmincon}. Let $\hat{\boldsymbol{\alpha}}$ be the solution of this problem. Then, the control node locations are found by solving
\begin{align}
& \min_{\boldsymbol{\pi}}  \sum_{i}^{N} | \pi_{i}-\alpha_{i} |, \text{subj. to}\;\; \sum_{l=1}^{N} \pi_{l} \le M_{\text{max}} , \;\pi_{l}\in \{0,1 \}.  \label{relaxedSolutionNLPs}  
\end{align} 
where $\alpha_{i}$ is the $i$th entry of $\boldsymbol{\alpha}$. By introducing slack variables, the problem \eqref{relaxedSolutionNLPs} can be easily transformed into an Integer Linear Program (ILP)~\cite{achterberg2009scip,belotti2013mixed}. We solve this ILP using the branch and bound method implemented in the MATLAB function \texttt{intlinprog}. Once this problem is solved, we compute the control sequence by solving the optimization problem \eqref{OptimizationProblem011}-\eqref{constraint011}.

\section{Numerical Results}
\label{numericalResults}
This section presents numerical experiments. All simulations are performed on a computer with 16GB RAM and Intel\textsuperscript{R} Core\textsuperscript{TM} i7-7500 processor.
The used codes are provided online in~\cite{haberControlCode2020}. In all simulation experiments we set $\mathbf{Q}_{i}=\m I$ for all $i$.

\subsubsection{Duffing Oscillator Networks}\label{resultsDuffing} 
\begin{figure}[t]
	\centering 
	\includegraphics[scale=0.48,trim=0mm 0mm 0mm 0mm ,clip=true]{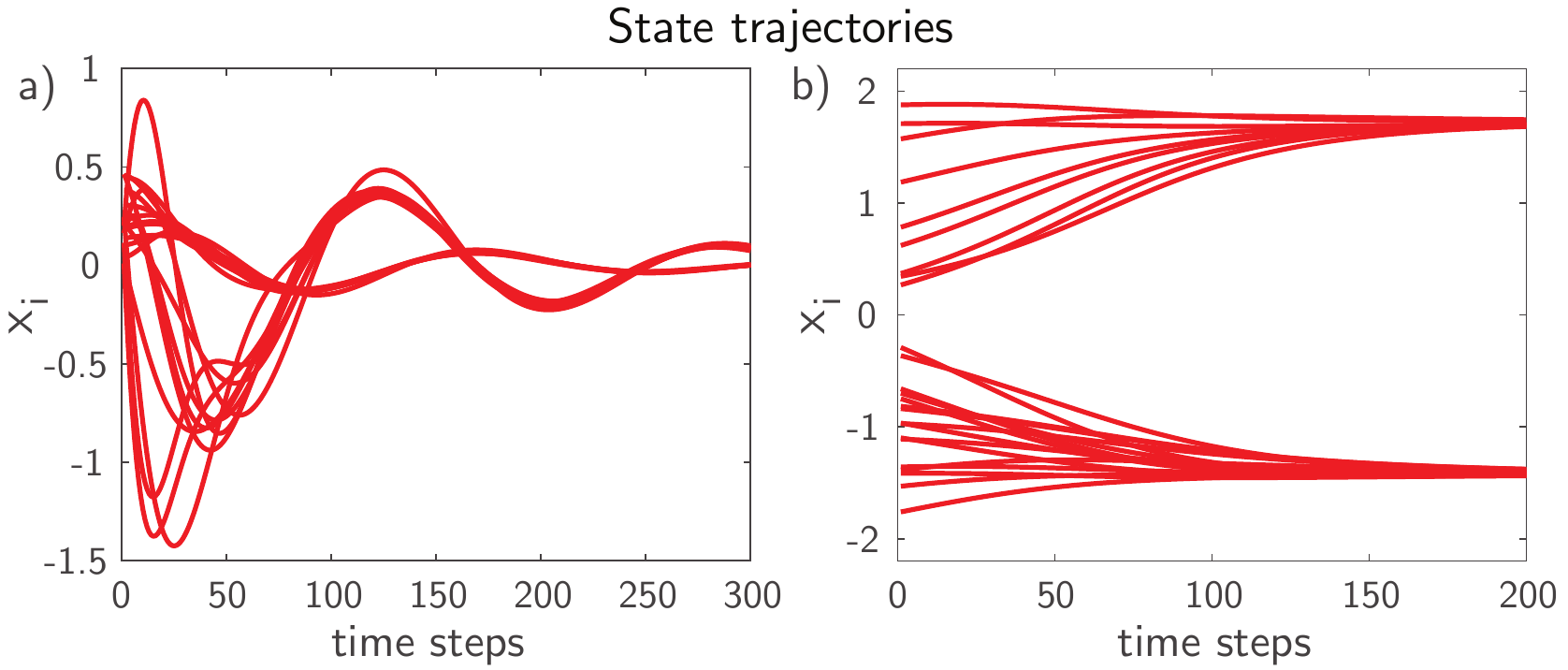}
	\vspace{-0.3cm}
	\caption{Uncontrolled responses of  (a) the Duffing oscillator network for $N=10$ (TI dynamics) and (b) the associative memory network for $N=25$ (FE dynamics). The results are generated for $h=10^{-2}$.}
	\label{fig:Graph1}
\end{figure}

\begin{figure}[t]\centering 
	\includegraphics[scale=0.46,trim=0mm 0mm 0mm 0mm ,clip=true]{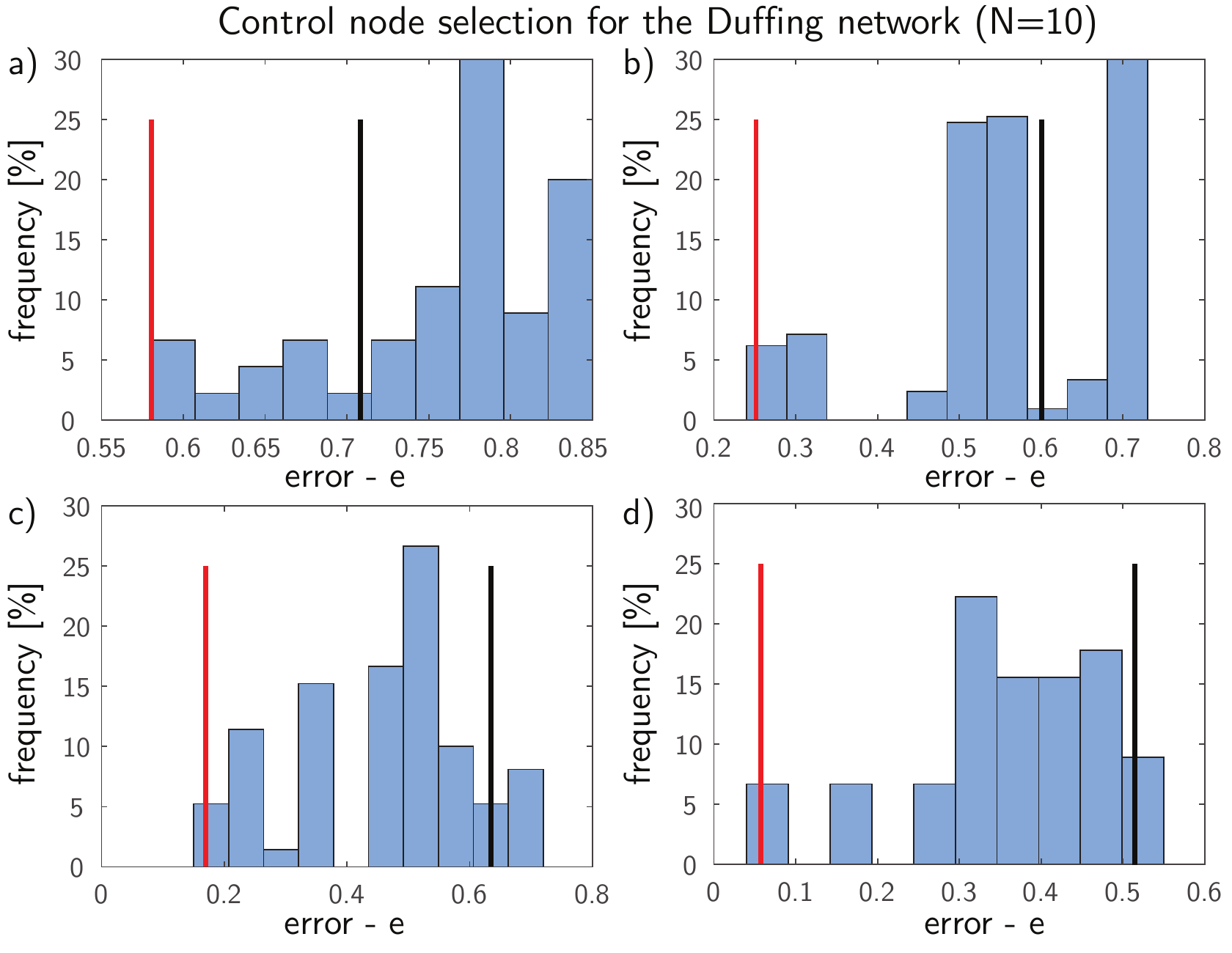}
		\vspace{-0.3cm}
	\caption{Control errors for the Duffing network with $10$ nodes. The red vertical line is the control error produced by Algorithm~\ref{algorithm1}. The black line is the control error produced by the method used for comparison. The histograms correspond to an exhaustive search for different fractions of control nodes. The results are generated for (a) $20\%$, (b) $40 \%$, (c) $60 \%$, and (d) $80 \%$ of fraction of control nodes.}\label{fig:Graph2}\end{figure}

\begin{figure}[t]
	\centering 
	\includegraphics[scale=0.46,trim=0mm 0mm 0mm 0mm ,clip=true]{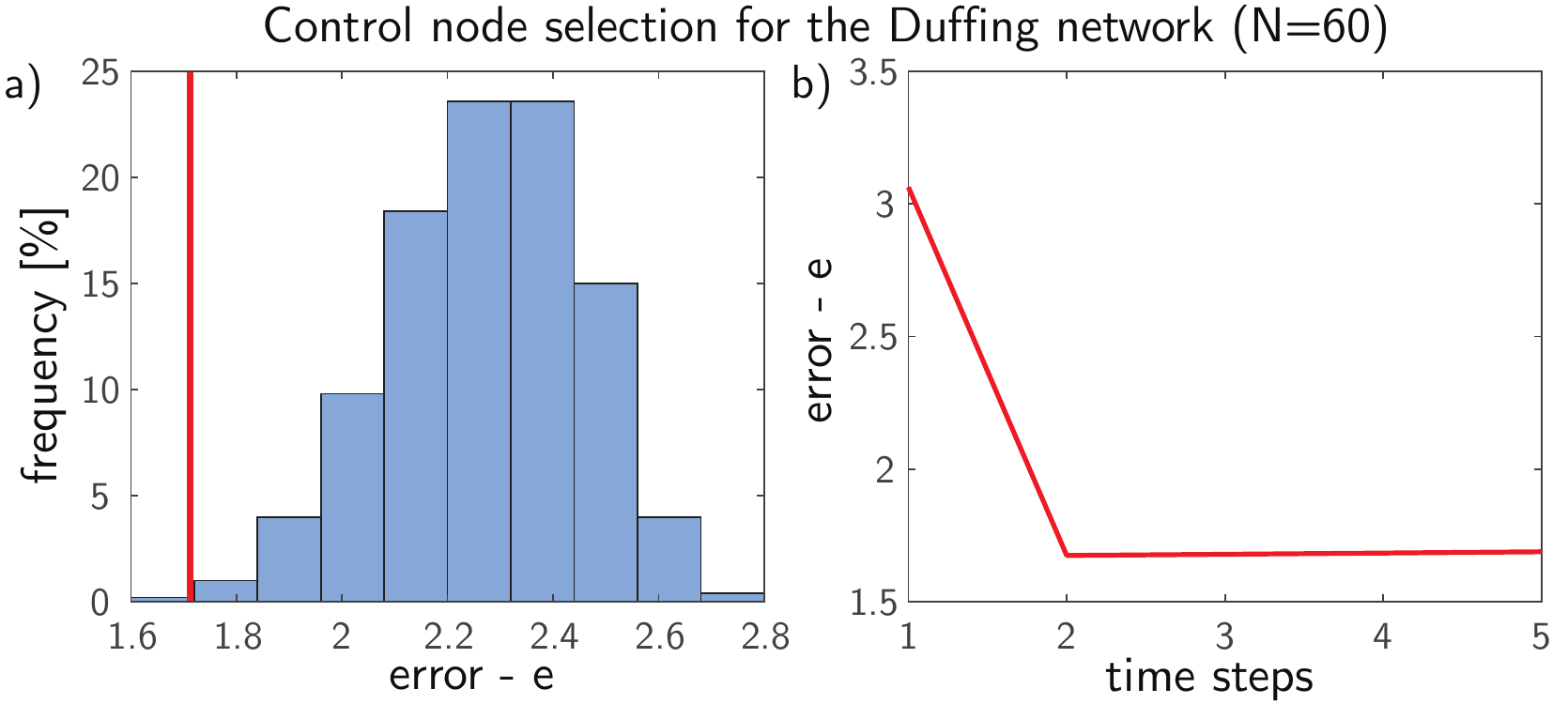}
		\vspace{-0.3cm}
	\caption{Control errors for the Duffing network with $60$ nodes. The histograms correspond to $500$ random selections of control nodes. (a) $47 \%$ control nodes. (b) The control error produced by Algorithm~\ref{algorithm1} as a function of control time steps.}
	\label{fig:Graph4}
\end{figure}

A large variety of physical systems, such as systems with geometric nonlinearities, electrical circuits, structural beams, cables, micromechanical structures, nanomechanical resonators, rotors, flight motor of an insect, etc., can be modeled by equations that closely resemble the governing equations of Duffing oscillators~\cite{kovacic2011duffing}. Duffing oscillators are characterized by a nonlinear spring stiffness $F_{s}=\alpha x - \beta x^{3}$, where $\alpha,\beta \in \mathbb{R}$ are the spring constants, $x$ is the spring displacement, and $F_{s}$ is the spring force (we assume a softening spring). We consider oscillator nodes connected via spring-damper connections 
\begin{align*}
& \dot{x}_{i1} = x_{i2},  \\
&\dot{x}_{i2}= -\alpha_{ii} x_{i1} +\beta_{ii}x_{i1}^{3}-\gamma_{ii}x_{i2}-\sum_{j\in \mathcal{N}(i)} \alpha_{ij}\big(x_{i1}-x_{j1}\big) \notag \\ &
+ \sum_{j\in \mathcal{N}(i)} \beta_{ij}\big(x_{i1}-x_{j1}\big)^{3}- \sum_{j\in \mathcal{N}(i)}\gamma_{ij}\big(x_{i2}-x_{j2}\big) +b_{i}u_{i },  
\end{align*}
where $x_{i1}$ and $x_{i2}$ are the position and velocity of the $i$-th oscillator, $\alpha_{ij}$ and $\beta_{ij}$ are the spring constants,  $\gamma_{ij}$ is damping, $b_{i}\in \{0,1 \}$ is the control parameter, $u_{i}$ is the control input,  $\mathcal{N}(i)$ denotes the set of nodes $j$ that are connected to the node $i$. The connection between oscillators is described by a  Geometric Random Graph (GRG) that is generated using the method and codes described in~\cite{taylor2009contest}. Such graphs tend to have larger diameters. The nodes are generated randomly on a unit square, and two nodes are connected if their spatial distance is below the radius of $\sqrt{1.44/N}$. The parameters $\alpha_{ij}$ are generated from a uniform distribution on the interval $[10,20]$, whereas the parameters $\beta_{ij},\gamma_{ij}$ are generated from a uniform distribution on the interval $[1,2]$. 
First, we consider a smaller Duffing oscillator network ($N=10$ nodes) for which we can perform an exhaustive search for controlled nodes. The network's uncontrolled response is shown in Fig.~\ref{fig:Graph1}(a).  
We use the following parameters $T=10$ and $h=10^{-4}$ and the TI discretization method. For this network, the FE method produces an unstable system response. We have simulated the uncontrolled dynamics using the MATLAB $\texttt{ode23s}$ function. The relative error between the TI method and the $\texttt{ode23s}$ solver is below $10^{-4}$. The entries of the desired vector are selected from a uniform distribution on the interval $[0,0.5]$. An initial network state for the MINO problem is computed as a steady-state (that approximates the equilibrium point) of the uncontrolled dynamics. For this purpose, the dynamics is simulated from a random state whose entries are generated from the uniform distribution on the interval $[ 0,0.5 ]$. In this way, we make the control problem more challenging, since we want to drive the network from a stable equilibrium point to a new unstable state. The control performance is quantified by computing the final control error: $e=\left\|\mathbf{x}_{D}-\mathbf{x}_{k=T} \right\|_{2}$. To test the method for fixed fractions of control nodes we replace the inequality in \eqref{constraint2} by equality. Vertical red lines in panels of Fig.~\ref{fig:Graph2} start from the error values on the horizontal axis that are obtained for the control sequence computed using Algorithm~\ref{algorithm1}. The black lines are errors produced by the method used for comparison that is summarized at the end of Section~\ref{controlProblemFormulation}. Histograms show control error distributions when control nodes are selected by exhaustive search (by exploring all the possible combinations for fixed fractions of controlled nodes). For each selection of control nodes (that determines the structure of matrix $\hat{\mathbf{B}}$) in the exhaustive search, the corresponding error is obtained for a control sequence computed by solving \eqref{problem011}. In this way, we can truly investigate and illustrate the main advantages of our method compared to exhaustive search and the method used for comparison. We can see that in most cases, the developed method generates optimal or almost optimal selections of control nodes. On the other hand, the method used for comparison does not produce as good results. 
Next, we generate a Duffing oscillator network with $N=60$ nodes ($Nn=120$) and all other parameters are unchanged. We test the method by keeping the inequality in~\eqref{constraint2}. The results analogous to the ones shown in Fig.~\ref{fig:Graph2} are shown in Fig.~\ref{fig:Graph4}. For brevity, we only show the results for $47\%$ of controlled nodes ($M_{\text{max}}=30$ and computed $M$ is $28$). 
Similarly to previously shown results, the results shown in Fig.~\ref{fig:Graph4} clearly demonstrate the good performance of the developed method.
\subsubsection{Associative Memory Networks}\label{resultsMemory}
 An associative memory network~\cite{cornelius2013realistic,nishikawa2004capacity} is used to memorize desired binary patterns (images of alphabet letters or signs). When a noisy image of a letter is set as the network's initial state, the network state should converge to the correct memorized letter. That is, the network should be able to recognize the correct letter that most closely resembles the one presented to it. From the dynamical system perspective,  memory networks encode memorized patterns as dynamically stable attractors~\cite{nishikawa2004capacity}.  An associative memory network consists of $N$ identical one-dimensional coupled oscillators~\cite{cornelius2013realistic,nishikawa2004capacity} 
\begin{align*}
& \dot{x}_{i}=\sum_{j=1}^{N} C_{ij} \text{sin}\big(x_{j}- x_{i}\big)+\frac{\varepsilon }{N} \sum_{j=1}^{N}\text{sin}2\big(x_{j}- x_{i}\big)+b_{i} u_{i}, 
\end{align*}
where $i=1,2,\ldots, N$, $x_{i},u_{i}\in \mathbb{R}$, $\varepsilon =0.8$ is the strength of the coupling term, $b_{i}\in \{ 0,1\}$, the coefficients are determined by Hebb's learning rule: $C_{ij}=(1/N) \sum_{\mu =1}^{p} \xi_{i}^{\mu}\xi_{j}^{\mu}$, where  $\boldsymbol{\xi}^{\mu}=\mathrm{col}\big( \xi_{1}^{\mu}, \xi_{2}^{\mu},\ldots, \xi_{N}^{\mu} \big)$, $\xi_{i}^{\mu}=\pm 1$, $\mu=1,2,\ldots, p$, and $p$ denotes the number of binary patterns to be memorized. The binary patterns $\boldsymbol{\xi}^{1}, \boldsymbol{\xi}^{2},\ldots, \boldsymbol{\xi}^{p}$ are a user choice. The goal is to memorize these patterns, such that when a perturbed pattern is given as an initial condition $\mathbf{x}(0)$, the network state should converge to the pattern that most closely resembles the correct pattern. The Hebb's learning rule ensures that $p$ desired binary patterns are coded as stable attractors of the system. 

\begin{figure}[t]\centering 
	\includegraphics[scale=0.46,trim=0mm 0mm 0mm 0mm ,clip=true]{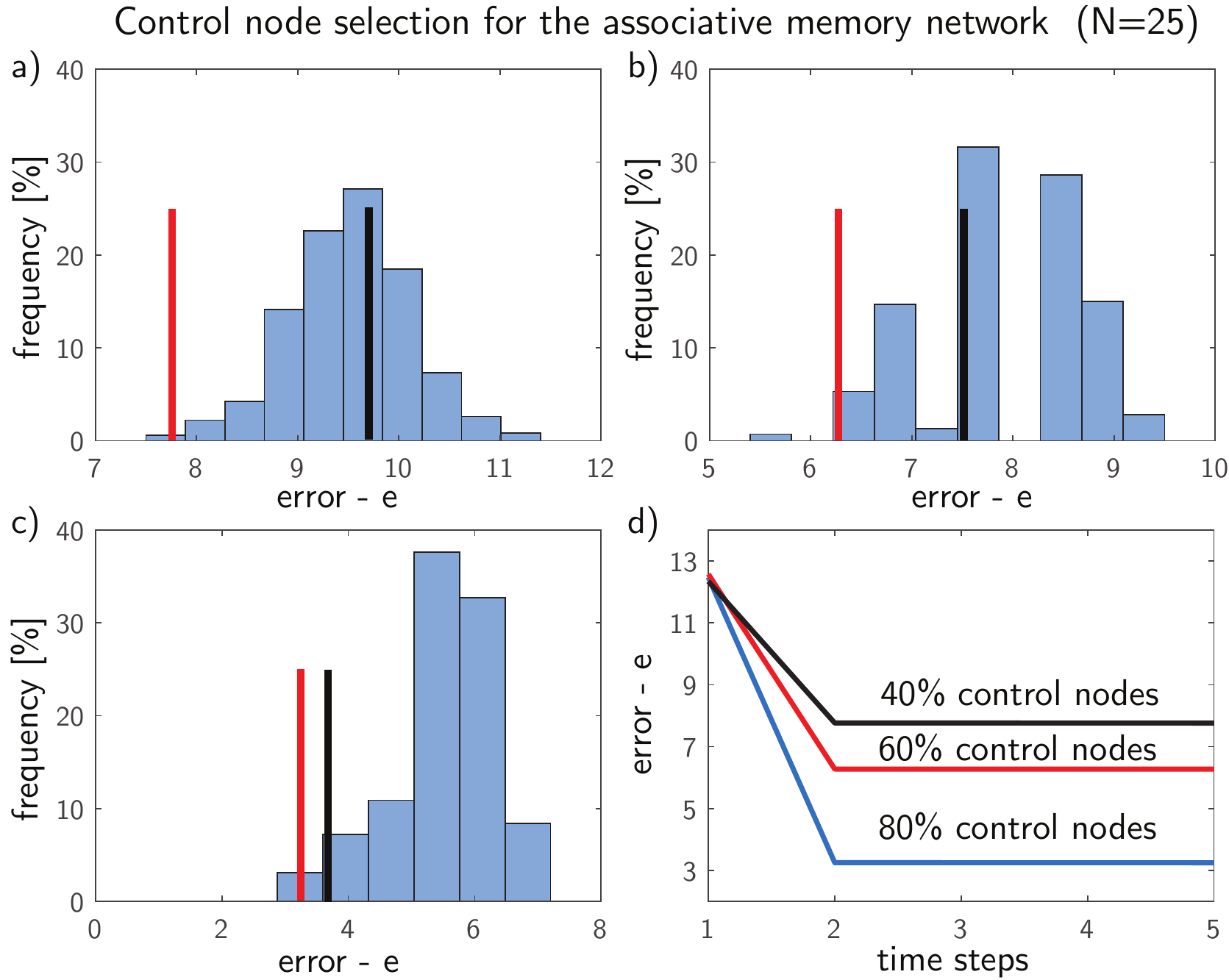}
		\vspace{-0.3cm}
		\caption{Control errors for the associative memory network with $25$ nodes.  (a)--(c) The red line is the control error produced by Algorithm~\ref{fig:Graph1}. The black line is the error produced by the method used for comparison. 		 The histograms correspond to $1000$ random selections of control nodes. (a) $40 \%$, (b) $60 \%$, and (c) $80 \%$ of fraction of control nodes. (d) Control errors produced by Algorithm~\ref{fig:Graph1} as functions of control time steps.}
	\label{fig:Graph3}
\end{figure}

 We select the following parameters $N=25$,  $T=10$, and $h=10^{-2}$. By comparing the simulated dynamics using the MATLAB $\texttt{ode45}$ function with the FE method, we concluded that the FE method for $h=10^{-2}$ is able to accurately simulate the dynamics (relative error bellow $10^{-3}$).  Consequently, we use the FE discretization method since the computational time of solving the MINO problem will be short (several minutes). The stored binary letters are defined on a $5\times 5$ mesh, and they are ``H", ``T", and ``L".  The network initial state is the ``H" letter that is perturbed by the normal Gaussian noise. Starting from this perturbed condition, the uncontrolled network will converge to the letter ``H".  The network's uncontrolled response is shown in Fig.~\ref{fig:Graph1}(b).  Starting from this initial state, our goal is to find the control nodes and the input sequence that will drive the network to the letter ``T". That is, we want to drive the network from the attractor of the letter ``H" to as close as possible to the letter ``T". To test the method for a fixed fraction of control nodes we replace the inequality in~\eqref{constraint2} by equality.  
 The results that are analogous to the results for the Duffing network are shown in Fig.~\ref{fig:Graph3}(a)--(c) except that instead of exhaustive search, we generate $1000$ random selections of control nodes (in this case, exhaustive search is computationally expensive). Figure~\ref{fig:Graph3}(d) shows the control error evolution for the first $5$ control steps. We see that after a single step the control error reaches the steady-state, and control inputs keep the network in the steady-state.

The results shown in Fig.~\ref{fig:Graph3} clearly demonstrate the excellent performance of the developed method. The proposed method is able to produce the final control error that is in most cases smaller than the error produced by randomly selecting the control nodes. Furthermore, we are able to reach the steady-state in a single control step.

\section{Conclusion, Limitation, and Future Work}
\label{conclusionsSection}
We have developed a control node selection method and tested it on two representative models of nonlinear networks. The simulation results demonstrate the good potential of the developed method. Some of the limitations of our approach are as follows. First, in the general case, the resulting MINO problem is non-convex and consequently, the computation of the optimal solution might be challenging for large-scale networks. Second, the proposed approach assumes an ideal model of the networks and does not take into account any disturbances that might come from modeling error, parameter uncertainty, and unknown inputs. Thirdly, the solution depends on initial and desired states, and possibly such a solution will change for different sets of states. In our future research---in addition to addressing the aforementioned limitations---we will explore different approaches for reducing the computational complexity and implementation efficiency (parallelization) of the method such that it can be used for large scale networks having hundreds and thousands of nodes. Furthermore, theoretical and numerical insights into the convergence and computational complexity of the developed method will also be investigated. Finally, we will generalize our problem formulation to feedback control scenarios and to the case when the number of control nodes is penalized.



\bibliographystyle{IEEEtran}

\end{document}

%% file: preamble.tex
\usepackage{epsfig,color,amsmath,cite}
\usepackage{amsmath}    
\IEEEoverridecommandlockouts
\usepackage{bm}
\usepackage{epstopdf}
\usepackage{amssymb}
\usepackage{url}
\let\labelindent\relax
\usepackage{enumitem}
\usepackage{multirow}
\usepackage{hhline}
\usepackage{booktabs}
\usepackage{threeparttable}
\usepackage{makecell}
\usepackage{xcolor}
\usepackage[linesnumbered,boxed,commentsnumbered,ruled,vlined,longend]{algorithm2e}
\usepackage{comment}
\setlist[itemize]{leftmargin=*}

\makeatother
\DeclareMathAlphabet\mathbfcal{OMS}{cmsy}{b}{n}



\usepackage{stackengine}

\newcommand{\m}{\boldsymbol}
\allowdisplaybreaks[4]
\usepackage[colorlinks = true,
linkcolor = blue,
urlcolor  = blue,
citecolor = blue,
anchorcolor = blue]{hyperref}

\usepackage[framemethod=TikZ]{mdframed}
\mdfdefinestyle{MyFrame}{%
	linecolor=black,
	outerlinewidth=1.25pt,
	roundcorner=1.25pt,
	innerrightmargin=5pt,
	innerleftmargin=5pt,}
	
\usepackage[noabbrev]{cleveref}

\usepackage{mathtools}

\DeclarePairedDelimiter\abs{\lvert}{\rvert}%
\DeclarePairedDelimiter\norm{\lVert}{\rVert}%

\makeatletter
\let\oldabs\abs
\def\abs{\@ifstar{\oldabs}{\oldabs*}}
\let\oldnorm\norm
\def\norm{\@ifstar{\oldnorm}{\oldnorm*}}
\makeatother

\usepackage[english]{babel}
\usepackage[utf8]{inputenc}
\usepackage[super]{nth}

\usepackage{graphicx}
\usepackage{float}
\usepackage[caption = false]{subfig}

%% file: ms_ieee.bbl
\begin{thebibliography}{10}
\providecommand{\url}[1]{#1}
\csname url@samestyle\endcsname
\providecommand{\newblock}{\relax}
\providecommand{\bibinfo}[2]{#2}
\providecommand{\BIBentrySTDinterwordspacing}{\spaceskip=0pt\relax}
\providecommand{\BIBentryALTinterwordstretchfactor}{4}
\providecommand{\BIBentryALTinterwordspacing}{\spaceskip=\fontdimen2\font plus
\BIBentryALTinterwordstretchfactor\fontdimen3\font minus
  \fontdimen4\font\relax}
\providecommand{\BIBforeignlanguage}[2]{{%
\expandafter\ifx\csname l@#1\endcsname\relax
\typeout{** WARNING: IEEEtran.bst: No hyphenation pattern has been}%
\typeout{** loaded for the language `#1'. Using the pattern for}%
\typeout{** the default language instead.}%
\else
\language=\csname l@#1\endcsname
\fi
#2}}
\providecommand{\BIBdecl}{\relax}
\BIBdecl

\bibitem{liu2011controllability}
Y.-Y. Liu, J.-J. Slotine, and A.-L. Barab{\'a}si, ``Controllability of complex
  networks,'' \emph{Nature}, vol. 473, no. 7346, pp. 167--173, 2011.

\bibitem{liu2013observability}
Y.-Y. Liu, J.-J. Slotine, and A.~Barab{\'a}si, ``Observability of complex
  systems,'' \emph{Proc. Natl. Acad. USA}, vol. 110, no.~7, pp. 2460--2465,
  2013.

\bibitem{cornelius2013realistic}
S.~P. Cornelius, W.~L. Kath, and A.~E. Motter, ``Realistic control of network
  dynamics,'' \emph{Nat. Commun.}, vol.~4, no.~1, pp. 1--9, 2013.

\bibitem{angulo2019theoretical}
M.~T. Angulo, C.~H. Moog, and Y.-Y. Liu, ``A theoretical framework for
  controlling complex microbial communities,'' \emph{Nat. Commun.}, vol.~10,
  no.~1, pp. 1--12, 2019.

\bibitem{haber2014subspace}
A.~Haber and M.~Verhaegen, ``Subspace identification of large-scale
  interconnected systems,'' \emph{{IEEE} Trans. Automat. Contr.}, vol.~59,
  no.~10, pp. 2754--2759, 2014.

\bibitem{haber2018sparsity}
------, ``Sparsity preserving optimal control of discretized pde systems,''
  \emph{{Comput. Methods Appl. Mech. Eng.}}, vol. 335, pp. 610--630, 2018.

\bibitem{reinschke1988multivariable}
K.~Reinschke, \emph{{Multivariable Control: A Graph-Theoretic Approach}}.\hskip
  1em plus 0.5em minus 0.4em\relax Springer, 1988, vol.~41.

\bibitem{pasqualetti2014controllability}
F.~Pasqualetti, S.~Zampieri, and F.~Bullo, ``Controllability metrics,
  limitations and algorithms for complex networks,'' \emph{IEEE Control Netw.
  Syst.}, vol.~1, no.~1, pp. 40--52, 2014.

\bibitem{haber2017state}
A.~Haber, F.~Molnar, and A.~E. Motter, ``State observation and sensor selection
  for nonlinear networks,'' \emph{IEEE Control Netw. Syst.}, vol.~5, no.~2, pp.
  694--708, 2017.

\bibitem{summers2015submodularity}
T.~H. Summers, F.~L. Cortesi, and J.~Lygeros, ``On submodularity and
  controllability in complex dynamical networks,'' \emph{IEEE Control Netw.
  Syst.}, vol.~3, no.~1, pp. 91--101, 2015.

\bibitem{letellier2018nonlinear}
C.~Letellier, I.~Sendi{\~n}a-Nadal, and L.~A. Aguirre, ``Nonlinear graph-based
  theory for dynamical network observability,'' \emph{Phys. Rev. E}, vol.~98,
  no.~2, p. 020303, 2018.

\bibitem{aguirre2018structural}
L.~A. Aguirre, L.~L. Portes, and C.~Letellier, ``Structural, dynamical and
  symbolic observability: From dynamical systems to networks,'' \emph{PLOS
  ONE}, vol.~13, no.~10, 2018.

\bibitem{qi2014optimal}
J.~Qi, K.~Sun, and W.~Kang, ``Optimal {PMU} placement for power system dynamic
  state estimation by using empirical observability {Gramian},'' \emph{IEEE
  Trans. Power Syst.}, vol.~30, no.~4, pp. 2041--2054, 2014.

\bibitem{nugroho2019algorithms}
S.~A. Nugroho, A.~F. Taha, N.~Gatsis, T.~H. Summers, and R.~Krishnan,
  ``Algorithms for joint sensor and control nodes selection in dynamic
  networks,'' \emph{Automatica}, vol. 106, pp. 124--133, 2019.

\bibitem{taha2018time}
A.~F. Taha, N.~Gatsis, T.~Summers, and S.~A. Nugroho, ``Time-varying sensor and
  actuator selection for uncertain cyber-physical systems,'' \emph{IEEE Control
  Netw. Syst.}, vol.~6, no.~2, pp. 750--762, 2018.

\bibitem{chang2018co}
C.-Y. Chang, S.~Mart{\'\i}nez, and J.~Cort{\'e}s, ``Co-optimization of control
  and actuator selection for cyber-physical systems,'' vol.~51, no.~23, 2018,
  pp. 118--123.

\bibitem{hao2019linear}
Y.~Hao, T.~Wang, G.~Li, and C.~Wen, ``Linear quadratic optimal control of
  time-invariant linear networks with selectable input matrix,'' \emph{(in
  press) IEEE Trans. Cybern.}, 2019.

\bibitem{taylor2016allocating}
J.~A. Taylor, N.~Luangsomboon, and D.~Fooladivanda, ``Allocating sensors and
  actuators via optimal estimation and control,'' \emph{IEEE Trans. Control
  Syst. Technol.}, vol.~25, no.~3, pp. 1060--1067, 2016.

\bibitem{nugroho2019sensor}
S.~A. {Nugroho} and A.~F. {Taha}, ``Sensor placement strategies for some
  classes of nonlinear dynamic systems via lyapunov theory,'' in \emph{2019
  IEEE 58th Conference on Decision and Control (CDC)}, 2019, pp. 4551--4556.

\bibitem{van2001review}
M.~Van De~Wal and B.~De~Jager, ``A review of methods for input/output
  selection,'' \emph{Automatica}, vol.~37, no.~4, pp. 487--510, 2001.

\bibitem{lou2003optimal}
Y.~Lou and P.~D. Christofides, ``Optimal actuator/sensor placement for
  nonlinear control of the {Kuramoto-Sivashinsky} equation,'' \emph{IEEE Trans.
  Control Syst. Technol.}, vol.~11, no.~5, pp. 737--745, 2003.

\bibitem{morris2010linear}
K.~Morris, ``Linear-quadratic optimal actuator location,'' \emph{{IEEE} Trans.
  Automat. Contr.}, vol.~56, no.~1, pp. 113--124, 2010.

\bibitem{edalatzadeh2019optimal2}
M.~S. Edalatzadeh and K.~A. Morris, ``Optimal controller and actuator design
  for nonlinear parabolic systems,'' \emph{arXiv:1910.03124}, 2019.

\bibitem{edalatzadeh2019optimal}
------, ``Optimal actuator design for semilinear systems,'' \emph{SIAM J.
  Control Optim.}, vol.~57, no.~4, pp. 2992--3020, 2019.

\bibitem{le2011algorithm}
S.~Le~Digabel, ``{Algorithm 909: NOMAD: Nonlinear optimization with the MADS
  algorithm},'' \emph{ACM Trans. Math. Softw.}, vol.~37, no.~4, pp. 1--15,
  2011.

\bibitem{belotti2013mixed}
P.~Belotti, C.~Kirches, S.~Leyffer, J.~Linderoth, J.~Luedtke, and A.~Mahajan,
  ``Mixed-integer nonlinear optimization,'' \emph{Acta Numerica}, vol.~22, pp.
  1--131, 2013.

\bibitem{haberControlCode2020}
\BIBentryALTinterwordspacing
A.~Haber, ``{Control Node Selection and Control Action Design for Nonlinear
  Networks and Systems}.'' [Online]. Available: \url{https://bit.ly/3fl9cuK}
\BIBentrySTDinterwordspacing

\bibitem{lars2011nonlinear}
L.~Gr{\"u}ne and J.~Pannek, \emph{Nonlinear Model Predictive Control : Theory
  and Algorithms}.\hskip 1em plus 0.5em minus 0.4em\relax Springer, 2011.

\bibitem{iserles2009first}
A.~Iserles, \emph{A first course in the numerical analysis of differential
  equations}.\hskip 1em plus 0.5em minus 0.4em\relax Cambridge university
  press, 2009, no.~44.

\bibitem{currie2012opti}
J.~Currie, D.~I. Wilson, N.~Sahinidis, and J.~Pinto, ``{OPTI: Lowering the
  barrier between open source optimizers and the industrial MATLAB user},''
  vol.~24, 2012, p.~32.

\bibitem{achterberg2009scip}
T.~Achterberg, ``{SCIP: solving constraint integer programs},'' \emph{Math.
  Program. Comput.}, vol.~1, no.~1, pp. 1--41, 2009.

\bibitem{sager2009reformulations}
S.~Sager, ``Reformulations and algorithms for the optimization of switching
  decisions in nonlinear optimal control,'' \emph{Journal of Process Control},
  vol.~19, no.~8, pp. 1238--1247, 2009.

\bibitem{sager2011combinatorial}
S.~Sager, M.~Jung, and C.~Kirches, ``Combinatorial integral approximation,''
  \emph{Mathematical Methods of Operations Research}, vol.~73, no.~3, p. 363,
  2011.

\bibitem{burger2019design}
A.~B{\"u}rger, C.~Zeile, A.~Altmann-Dieses, S.~Sager, and M.~Diehl, ``Design,
  implementation and simulation of an {MPC} algorithm for switched nonlinear
  systems under combinatorial constraints,'' \emph{Journal of Process Control},
  vol.~81, pp. 15--30, 2019.

\bibitem{kovacic2011duffing}
I.~Kovacic and M.~J. Brennan, \emph{{The Duffing Equation: Nonlinear
  Oscillators and Their Behaviour}}.\hskip 1em plus 0.5em minus 0.4em\relax
  John Wiley \& Sons, 2011.

\bibitem{taylor2009contest}
A.~Taylor and D.~J. Higham, ``{CONTEST: A controllable test matrix toolbox for
  MATLAB},'' \emph{ACM Trans. Math. Softw.}, vol.~35, no.~4, pp. 1--17, 2009.

\bibitem{nishikawa2004capacity}
T.~Nishikawa, Y.-C. Lai, and F.~C. Hoppensteadt, ``Capacity of oscillatory
  associative-memory networks with error-free retrieval,'' \emph{Phys. Rev.
  Lett.}, vol.~92, no.~10, p. 108101, 2004.

\end{thebibliography}
